# Dynamical Response of Deformable Microchannels under Pressure-Driven Flow of Aqueous Polymer Solutions

Sampad Laha[1], Siddhartha Mukherjee [2], and Suman Chakraborty[1]*

[1]Department of Mechanical Engineering, Indian Institute of Technology Kharagpur, Kharagpur, West Bengal, 721302, India.

[2]Department of Chemical Engineering, Indian Institute of Technology Guwahati, Guwahati, Assam, 781039, India.

*Email: suman@mech.iitkgp.ac.in

## Abstract

Microfluidic channels are integral to biomedical technology and process engineering, offering versatility in handling fluids with complex properties, often a combination of viscous and elastic attributes. Despite significant advancements in understanding small-scale fluid-structure interactions, however, experimental insights on the flow of complex fluids in deformable microchannels remain limited. Here, we present controlled experiments using polymer solutions as model viscoelastic fluids to examine the effects of polymer concentration on the elasto-mechanical characteristics of slender cylindrical microchannels. The findings indicate significant differences in fluid-structure interactions between dilute and semi-dilute polymer solutions with varying molecular weights. At higher polymer concentrations, these interactions intensify, leading to reduced pressure drops in high-shear regions and increased pressure drops in low-shear areas, linked to local wall deformation. The increased elasticity of higher concentration solutions further enhances local deformation, disrupts flow, and dissipates energy, resulting in a non-linear rise in pressure drop. This behaviour is aggravated by the solution's increased apparent viscosity due to the entangled polymer network. A theoretical model of flow-induced deformation is also developed, accounting for polymer chain extensibility. These insights highlight the importance of polymer constitution in optimizing the flow characteristics, advancing the development of adaptive microfluidic devices in biological and industrial applications for optimal performance.

**Impact Statement**

The interaction between polymer solutions with complex rheology and deformable microchannels is crucial for both biological systems and industrial processes, as well as for advancing adaptive microfluidic devices. Deformable microchannels, whose geometry can be modified by external stimuli, offer a platform to optimize flow properties and enhance polymer processing, such as in polymer extrusion or printing. This dynamic adaptability is particularly useful for controlling product morphology in industries involving advanced materials and additive manufacturing. Our experimental findings on the pressurized flow of polymer solutions through deformable microchannels provide critical insights into the complex interactions between polymeric behaviour and channel deformability, an area previously unexplored. Complemented by theoretical analysis, our research allows for a comparative understanding of the elasto-mechanical response of microfluidic systems across different complex fluids, offering a foundation for improving the design of microfluidic devices in real-world applications.

**Keywords:** deformability; fluid-structure interaction; viscoelastic fluids; polymer concentration; molecular weight; microchannel.

# 1. Introduction

Microfluidic channels often exhibit flexibility and interact with the flowing liquid (Chakraborty *et al.* 2012; Duprat C 2016; Grotberg & Jensen 2004; Hardy *et al.* 2009; Sandberg LB, Gray WR 1977; Skotheim & Mahadevan 2004, 2005; Xia & Whitesides 1998; Raj *et al.* 2017; Christov *et al.* 2018). These deformable conduits may provide a more accurate simulation of several real-world scenarios as compared to their rigid counterparts (Dendukuri *et al.* 2007; Jeong *et al.* 2005; Karan *et al.* 2020; Panda *et al.* 2009; Seker *et al.* 2009; Singh *et al.* 2015; Wang *et al.* 2022a). Consequently, insights obtained from their dynamic characteristics may provide essential cues for intricate control of complex fluids flowing therein, which is essential for high-precision applications, such as manipulating small volumes of liquids and directing them to targeted sites (Anand *et al.* 2019; Boyko *et al.* 2017; Chun *et al.* 2024). In the oil industry, understanding how complex fluids behave in such channels can optimize enhanced oil recovery techniques, such as polymer flooding, by improving reservoir simulation models (Mohsenatabar Firozjaii & Saghafi 2020; Rellegadla *et al.* 2017; Speight 2013). In polymer processing, these studies lead to better control over material properties, more efficient manufacturing, and higher-quality products. Deformable microchannels are also crucial in advancing wearable technologies, enabling the development of flexible electronics and soft robotics that conform to the body, ensuring reliable performance(Gao & Ren 2021; Kim *et al.* 2018; Xu *et al.* 2014). In the biomedical realm, microvasculature-on-a-chip technologies (Benam *et al.* 2016; Bhattacharya *et al.* 2022; Huh 2015; Priyadarshani *et al.* 2021) benefit by accurately simulating blood vessel dynamics, which is essential for studying physiological processes and drug interactions. These studies may facilitate more personalized and effective treatments than ever before, by replicating patient-specific vascular characteristics, leading to improved drug delivery systems and therapeutic strategies (Palasantzas *et al.* 2023; Wang *et al.* 2023). Overall, flow through deformable microchannels is fundamental to innovations in energy, manufacturing, healthcare, and wearable technologies (Chung *et al.* 2019; M. *et al.* 2018; Zhang *et al.* 2020). In such rapidly evolving arena, the medium - whether a biological fluid like blood or an industrial fluid like polymer - may display unique stress-responsive behaviours that combine viscous and elastic properties, significantly differing from purely viscous interactions (Brust *et al.* 2013; Kawata *et al.* 2000; Thurston 1972).

Polymer solutions, which are the primary focus of this work, belong to a class of viscoelastic fluids that interact uniquely with the mechanical behaviour of the channel walls, depending on the fluid's composition. In fluid-structure interaction problems, when fluids exert pressure on the microchannel walls, the walls tend to deform (Anand *et al.* 2019; Anand & Christov 2020; Boyko *et al.* 2017, 2019, 2020, 2022; Chakraborty & Chakraborty 2010, 2011; Christov 2021; Christov *et al.* 2018; Elbaz & Gat 2014, 2016). This deformation, in turn, alters the flow field and eventually leads to a feedback loop - a hallmark of fluid-structure interaction phenomenon (Inamdar *et al.* 2020; Martínez-Calvo *et al.* 2020; Mukherjee *et al.* 2022; Venkatesh *et al.* 2022; Wang *et al.* 2022; Wang & Christov 2019; Zhang *et al.* 2020). Now, for the flow of polymer solutions in deformable channels, the fluid has its own elastic characteristics distinct from the wall's elastic response that can lead to non-uniform deformation of the microchannel. The altered channel shape further influences the flow, resulting in a situation where the pressure drops required to maintain a certain flow rate differ significantly from those in rigid channels as well as deformable channels conveying purely viscous fluids. This is due to a confluence of the viscoelastic nature of the fluid and the extent of wall deformation, potentially leading to non-linear scaling of flow rate with pressure(Boyko & Christov 2023; Ramos-Arzola & Bautista 2021). Moreover, the flow of viscoelastic fluids in deformable channels may exhibit relaxation dynamics as the fluid and channel walls interact, which is particularly relevant in microfluidic pumps and biological systems (Dendukuri *et al.* 2007; Hardy *et al.* 2009; Hosokawa *et al.* 2002; Iyer *et al.* 2015; Jeong *et al.* 2005; Kartalov *et al.* 2007; Mukherjee *et al.* 2013; Panda *et al.* 2009; Seker *et al.* 2009; Singh *et al.* 2015; Roy *et al.* 2020; Sumets *et al.* 2018).

Despite recent theoretical efforts to analyse the flow of non-Newtonian fluids in deformable microchannels ( Boyko *et al.* 2017; Nahar *et al.* 2019; Raj & Sen 2016; Raj M *et al.* 2018; Tanner *et al.* 2012; Boyko & Christov 2023; Ramos-Arzola & Bautista 2021), experimental studies in this area remain limited. Recently, experiments have been conducted to explore elasto-inertial instability and the transition from laminar to turbulent flow inside deformable tubes (Chandra *et al.* 2019). However, these instabilities typically occur at high Reynolds numbers, whereas many microfluidic applications involve low Reynolds number flows, for which experimental insights into the interaction between viscoelastic fluids and their confining boundaries are relatively scarce. A critical factor in these interactions is the concentration of the polymer solution as a model representative of its viscoelastic attribute, which significantly influences the pressure drop and flow behaviour. As the polymer concentration increases, the apparent viscosity of the solution also rises (Del Giudice *et al.* 2015). This is because the large polymer molecules create entanglements and interactions within the fluid, resisting the flow (R.G.Larson 1999). Additionally, many polymer solutions exhibit shear-thinning behaviour, where viscosity decreases with increasing shear rate (Boyko & Christov 2023). At higher polymer concentrations, this effect becomes more pronounced, potentially reducing the pressure drop in high-shear regions but increasing it in low-shear areas, especially if local variations in channel wall deformation are involved. Polymers can also introduce elastic effects into the flow, particularly at high concentrations, which may lead to flow instabilities, oscillations, or even flow-induced vibrations under certain conditions (Dey *et al.* 2020; Wu *et al.* 2021). The resulting deformation of the channel can cause localized irregularities, such as bulges or constrictions, disrupting flow, increasing energy dissipation, and potentially leading to higher pressure drops. These irregularities can also create localized areas of low or high pressure, further complicating the overall pressure drop behaviour along the channel. In applications like enhanced oil recovery, where polymers are used to improve the sweep efficiency of water floods, optimizing polymer concentration is essential to balance viscosity increase with drag reduction, achieving the desired pressure drop and flow characteristics (Kakati *et al.* 2022; Speight 2013). Similarly, in microfluidic devices, precise control of polymer concentration is necessary to maintain desired flow profiles without causing excessive pressure drops. However, the complex interaction between polymer concentration and the elastic response of the channel walls can result in dynamic features that are far from being trivial, as compared to the dynamical responses of purely viscous fluids in rigid or deformable microchannels.

Here, we present experiments that investigate the effect of viscoelastic rheology on the elasto-mechanical characteristics of slender cylindrical microchannels. Towards that, we report on experiments conducted with polymer solutions of varying compositions in both dilute and semi-dilute regimes to unravel the impact of its molecular weight and concentration on the deformation landscape. To explain our experimental findings, we present a theoretical model based on the Finitely Extensible Nonlinear Elastic-Peterlin (FENE-P) constitutive behaviour (Peterlin 1961). This model accounts for the non-linear elasticity of polymer chains that enables their stretching only to a finite length. As a polymer chain approaches its maximum extensibility, the force required for further stretching increases significantly, capturing the realistic behaviour of polymers under flow conditions. The Peterlin approximation, which assumes that the average stretch of the polymer chain can be represented by a scalar function of the conformation tensor, simplifies the underlying conceptual depiction without sacrificing the essential physics of interest (Oliveira 2002). The model thus modifies the Oldroyd-B constitutive equation by incorporating a factor that accounts for the finite extensibility of polymer chains. It can thus effectively capture the shear-thinning behaviour commonly observed in polymer solutions in real-life scenarios, where the viscosity decreases with increasing shear rate due to the alignment and stretching of polymer chains in the direction of flow. The model also accounts for the development of normal stress differences, which can lead to phenomena such as rod climbing or die swell often observed in experiments (Batchelor *et al.* 1973; More *et al.* 2023; Weissenberg 1947). This quantitative mapping of the interplay between viscoelastic rheology and structural compliance has significant potential to advance the development of design strategies for flexible microfluidic flow-actuating and manipulating devices using polymeric fluids. These findings may also bear critical implications for bioengineered in-vitro devices, flexible electronics, enhanced oil recovery, chemical processing, and beyond.

## 2. Materials and Methods

### 2.1. Microchannel fabrication

The fabrication of circular microchannels has been done using the "pull-out" soft lithography method as reported previously using polydimethylsiloxane (PDMS) (Laha *et al.* 2024; Raj M. *et al.* 2018) . In order to make the channels, a lumber puncture (LP) needle no. 22 (Comet India) was used as the negative replicating mould. The LP needle set usually comes with a solid inner needle which fits coaxially into a hollow outer needle case. The LP 22 set corresponds to an inner needle of diameter of ~ 340 µm. In order to make the casting chamber, two identical holes have been drilled at the diametrically opposite sides of a 35 mm plastic petri dish and 5 mm length pieces of the outer needle were fixed onto these holes using Araldite. Thereafter, the inner LP 22 needle is inserted into the petri dish over which the PDMS (Sylgard 184, Dow Corning, USA) and crosslinker mixture could be poured and cured. The elastomer and crosslinker are mixed thoroughly in 10:1, 20: 1 and 30:1 ratio (w/w), degassed in a vacuum desiccator, poured over the mould and kept at 85 °C for 14 hours for curing inside a hot air oven. The 10:1 polymer mixture corresponds to a rigid non-deformable channel while the 20:1 and 30:1 mixture both produce deformable microchannels (refer to Table S1 in Supplementary material (SM) for the elastic properties of the different PDMS samples, Karan *et al.* 2020; Raj M. *et al.* 2018). Once cured properly, the needle is gently pulled out of the petri dish leading to the formation of a microchannel of cylindrical cross section with roughly identical diameter (~ 338.7 ± 3.43) as that of the needle. The outer needle extensions of the petri dish mould could not be used as fluidic connectors because of the repeated leakage occurring at the junction (inside the petri dish mould) between the inner end of the needles and the microchannel at high flow rates. This occurs primarily because of the mismatch in the diameters of the channel and that of the outer needle connectors corresponding to the LP 22 needle. Hence, in order to solve this issue, we scooped out the cured PDMS block gently from the petri dish mould and inserted 5 mm connectors made of the outer casing of the LP 24 needle (inner needle diameter ~ 240 µm), which closely matches the microchannel diameter. After inserting the connectors from both ends, the useable length of the microchannel visible for imaging was ~ 2.7 mm. Finally, the inlet and outlet junctions of the PDMS channel block are sealed with Araldite in order to ensure leakage-free operations.

## 2.2. Sample preparation

De-ionized water (DI water, Milli-Q, Millipore, India) is used as the Newtonian solvent. To analyse the effect of the viscoelastic fluid rheology on the deformation characteristics, viscoelastic fluids have been prepared first using aqueous solutions of three different polymers namely polyethylene oxide (PEO, molecular weight: 4000000 g/mol), polyvinyl alcohol (PVA, molecular weight: 115000 g/mol), and polyethylene glycol (PEG, molecular weight: 10000 g/mol). The concentration of the polymers in DI water is varied to prepare solutions in dilute and semi-dilute unentangled regimes. Before preparing the solutions, the cross-over concentration $c^*$ is calculated first for all three polymers which are 0.071% (w/vol), 1.06% (w/vol), and 5.3% (w/vol) for PEO, PVA, and PEG, respectively (Tirtaatmadja *et al.* 2006; Varma *et al.* 2020). For PEO, we have prepared four different concentrations $0.5c^*$, $c^*$, $2c^*$, and $4c^*$, corresponding to both dilute as well as semi-dilute regimes [Refer to Table S2 in SM for viscosity and relaxation times for different PEO concentrations]. To realize the alteration in the deformation characteristics for low molecular weight polymers, we have chosen PVA and PEG for which concentrations corresponding to $c^*$ (1.06% w/vol and 5.3% w/vol, respectively) values are prepared. The stirring time for preparation of PEO and PEG solutions in DI water are 24 hours and 6 hours, respectively while the solutions are kept at $40^{\circ}$C (Varma *et al.* 2020). For preparing the PVA solution, the stirring time is initially 0.5 hour at $90^{\circ}$C followed by stirring of 6 hours at $40^{\circ}$C (Varma *et al.* 2020).

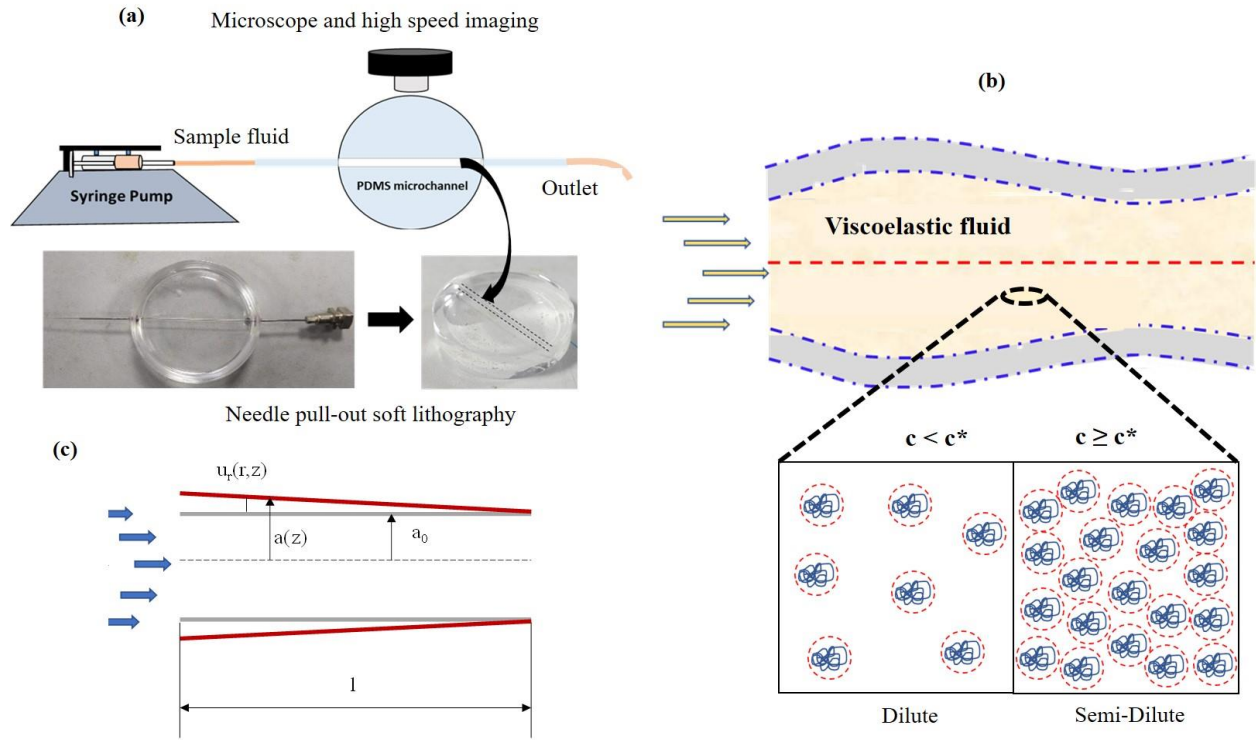

**Figure 1.** Schematic for the experimental setup and model development. (a) shows the schematic of the various instruments used for experiments including the microscope, syringe pump and microchannel. The needle pull-out soft lithography technique is used for fabrication as highlighted by the real-time images of the microchannel fabrication setup. (b) shows a schematic of a viscoelastic fluid flowing through a deformable channel, highlighting two different polymer concertation regimes: dilute and semi-dilute (c) shows a model schematic of axisymmetric channel deformation for theoretical calculations.

*2.3. Flow setup, Imaging and post-processing*

Figure 1 shows the schematic of the experimental setup used in this study. A syringe pump (Harvard PHD 2000) was used for a steady flow of liquid through the microfluidic channel. Teflon tubes served as the connectors for the inlet and outlet ports of the channel. The drainage tube connected to the outlet port was 45 mm long and 500 μm in diameter. The diameter of the deformable channels was measured under a wide range of flow rates starting from 50 μL/mins to 1000 μL/mins using an inverted optical microscope (OLYMPUS, IX71) in bright-field mode. A propriety image capturing software (ProgRes CapturePro 2.9.0.1) was used for recording the images of the deformed channels which were later processed in ImageJ software for the actual measurements. The channel diameter was measured near the inlet at a distance of ~ 1.7 mm downstream from the inlet port to avoid any deformation bias due to connector integration at the inlet. Sample microscopic images of microchannel deformation at different flow rates have been provided in Figure S1 of the SM. All the measurements have been taken after the flow and channel deformation reached steady state. The deformation measurements for each of the flow rates has been repeated three times to ensure the repeatability of the experimental data.

## 3. Theoretical overview

In this section, we provide the theoretical foundation of the present problem. We consider steady, laminar, and incompressible flow of a viscoelastic fluid through a deformable cylindrical microchannel (of length $l$ and radius $a_0$ of undeformed configuration) driven by a pressure-gradient. The model schematic of the problem is depicted in Figure 1(c). We have chosen the cylindrical coordinate system, in which $z$ is the axial coordinate and $r$ is the radial coordinate. With the circumferential symmetry assumption (i.e., '$\theta$' symmetry in which $\theta$ is the azimuthal coordinate), this analysis essentially simplifies into a two-dimensional problem. For incorporating the complex rheological characteristics of the viscoelastic fluid, the FENE-P constitutive model is considered. This necessitates the interplay between the fluid rheology and the elasticity of the confining boundary which in turn influences the resulting channel deformation. Thus, in this problem, we have both (i) the fluid domain as well as (ii) the solid domain. We first focus on the solid domain where the objective is to determine the elasto-mechanical response of the microchannel from the principles of solid mechanics. For this, the governing equation is the elastostatic equilibrium equation (Karan *et al.* 2020; Raj M *et al.* 2018; Wang *et al.* 2022b):

$$\nabla \cdot \sigma = 0 \qquad (3.1)$$

which governs the stress field of the solid domain. In equation (3.1), $\sigma$ is the Cauchy-Green stress tensor for the solid domain, i.e., where $\sigma = G(\nabla \boldsymbol{d} + \nabla \boldsymbol{d}^T) + \lambda' \nabla \cdot \boldsymbol{d}[\boldsymbol{I}]$ with $G$ and $\lambda'$ being the Lamé coefficients for an isotropic, linearly elastic solid. Here, '$\boldsymbol{d}$' is the displacement field vector, and $\boldsymbol{I}$ is the identity matrix. We here assume that the thickness of the deformable layer is very small compared to the axial length scale. Also, the degree of deformation occurring here is considered to be small enough so that we can stay in the linear elastic regime (Wang *et al.* 2022). In terms of the stress components, we now rewrite equation (3.1) as

$$\frac{1}{r}\frac{d}{dr}(r\,\sigma_{rr}) - \frac{\sigma_{\theta\theta}}{r} = 0 \qquad (3.2)$$

where the expressions of the stresses $\sigma_{rr}$ and $\sigma_{\theta\theta}$ are as follows: $\sigma_{rr} = \frac{\lambda'}{r}\frac{\partial}{\partial r}(r\,u_r) + 2G\frac{\partial u_r}{\partial r}$ and $\sigma_{\theta\theta} = \frac{\lambda'}{r}\frac{\partial}{\partial r}(r\,u_r) + 2G\frac{u_r}{r}$. Here, $u_r$ is the radial deformation of the solid domain and hence, the deformed radius can be defined as $a(z) = a_0 + u_r(z)$ with $a_0$ being the radius of the undeformed microchannel. We substitute these expressions of $\sigma_{rr}$ and $\sigma_{\theta\theta}$ in equation (3.2) to arrive at the following governing differential equation for estimating the displacement in the solid domain:

$$\frac{1}{r}\frac{\partial^2}{\partial r^2}(r\,u_r) - \frac{1}{r^2}\frac{\partial}{\partial r}(r\,u_r) = 0 \qquad (3.3)$$

We next apply the boundary conditions; the first one is the traction balance condition at the fluid-deformable layer interface (i.e., $\sigma_{rr}(a_0, z) = -p(z)$ where $\sigma_{rr}$ is the normal stress component corresponding to the solid domain and $p(z)$ is the pressure distribution corresponding to the fluid domain. The other one is the vanishing stress field away from the solid-fluid interface, i.e., $\lim_{r \to \infty} \sigma_{rr}(r, z) = 0$ (Wang *et al.* 2022b). The resulting displacement field of the solid domain is

$$u_r(r, z) = \frac{a_0^2}{2\,G\,r} p(z) \qquad (3.4)$$

and the deformed radius becomes $a(z) = a_0 \left\{1 + \frac{p(z)}{2\,G}\right\}$ (Wang *et al.* 2022b). The deformed radius involves contributions from both solid and fluid domains and thus, the flow dynamics become coupled with the deformation in the solid domain.

Now we shift our attention on the fluid domain. With the consideration of the slenderness of the microchannel (i.e., the axial length scale is much larger compared to the channel radius), we use the classical lubrication approximation which simplifies the governing equations considerably. For the steady, laminar, and incompressible flow, the continuity equation reads as $\frac{1}{r}\frac{\partial(rv_r)}{\partial r} + \frac{\partial v_z}{\partial z} = 0$. Now, the *r*-component of the Cauchy momentum equation gives us $0 = -\frac{\partial p}{\partial r}$ which implies $p \neq f(r)$ and function of *z* only. Using this, the *z*-component of the momentum equation can be written as

$$0 = -\frac{dp}{dz} + \frac{1}{r}\frac{\partial}{\partial r}(r\,\tau_{zr}) \qquad (3.5)$$

The next step is to incorporate the viscoelastic fluid rheology in the momentum equation. Typically, aqueous polymer solutions show viscoelastic characteristics the rheological behaviour of which is represented here by the FENE-P (i.e., Finitely Extensible Non-linear Elastic model with Peterlin's closure) constitutive model (Bird RB, Armstrong RC 1987; Oliveira 2002; Peterlin 1961)

$$f\,\boldsymbol{\tau} + \lambda \stackrel{\nabla}{\boldsymbol{\tau}} = 2\,b\,\mu\,\boldsymbol{D} \qquad (3.6)$$

The incorporation of the FENE-P constitutive model here not only considers the primary viscoelastic features like the shear-thinning effect, finite elasticity, normal stress difference, etc. but also captures the finite extensibility of polymer chain. Thus, it circumvents the limitation of the Oldroyd-B constitutive model. In this context it is important to mention that, the Oldroyd-B model was previously employed to examine the influence of viscoelastic fluid rheology on the fluid-structure interaction (Boyko & Christov 2023), where the degree of viscoelasticity was integrated to the system through an asymptotic approach using Deborah number (*De*) as the perturbation parameter (which represents the relative strength of elastic effect compared to viscous effect). While the asymptotic approach has its own limitation, analyzing the deformation characteristics by artificially changing the value of Deborah number might not prove to be a wise proposition, since in experimental scenarios, both the elastic and viscous behavior of the viscoelastic polymer solutions is strongly regulated by the polymer concentration. Here, we have not only widened the applicability of the developed fluid-structure interaction model by utilizing the FENE-P constitutive behavior, but also represented the fluid rheological parameters in a way that is realizable in experimental scenarios.

In equation (3.6), $\boldsymbol{\tau}$ is the viscous stress tensor, $\boldsymbol{D}$ is the rate of deformation tensor $\boldsymbol{D} = \frac{1}{2}(\nabla \boldsymbol{v} + \nabla \boldsymbol{v}^T)$, $\overset{\nabla}{\boldsymbol{\tau}}$ is the upper convective derivative $\overset{\nabla}{\boldsymbol{\tau}} = \frac{D\boldsymbol{\tau}}{Dt} - \nabla \boldsymbol{v}^T \cdot \boldsymbol{\tau} - \boldsymbol{\tau} \cdot \nabla \boldsymbol{v}$ with $\boldsymbol{v}$ being the velocity field with components $v_z$ and $v_r$, respectively. $\mu$ and $\lambda$ are the fluid viscosity and fluid relaxation time, respectively. Here, $b$ is an additional rheological parameter defined as $b = \frac{L^2}{L^2-3}$ with $L^2$ being the fluid extensibility parameter defined as the square of the ratio between the maximum and the equilibrium lengths of spring (Oliveira 2002). Here, $f$ is the stress coefficient function which takes the form $f = 1 + \frac{3b + \frac{\lambda}{\mu} tr\,\boldsymbol{\tau}}{L^2}$ where $tr\,\boldsymbol{\tau}$ is the trace of stress tensor, i.e., $tr\,\boldsymbol{\tau} = (\tau_{rr} + \tau_{zz})$.

With the assumptions mentioned earlier for the flow field, the stress components are simplified considerably which satisfy the following three equations

$$f\,\tau_{zz} = 2\lambda \frac{\partial v_z}{\partial r} \tau_{zr} \tag{3.7}$$

$$f\,\tau_{rr} = 0 \tag{3.8}$$

$$f\,\tau_{zr} = b\mu \frac{\partial v_z}{\partial r} + \lambda \frac{\partial v_z}{\partial r} \tau_{rr} \tag{3.9}$$

From equation (3.8), we can conclude that $\tau_{rr}$ must be equal to zero since the stress coefficient function $f$ is non-zero. Substituting $\tau_{rr} = 0$ in equation (3.9) yields

$$f\,\tau_{zr} = b\mu \frac{\partial v_z}{\partial r} \tag{3.10}$$

Dividing equation (3.7) by equation (3.10) gives us $\tau_{zz} = \frac{2\lambda}{b\mu} \tau_{zr}^2$. Substituting this relationship between the normal stress ($\tau_{zz}$) and tangential stress ($\tau_{zr}$) in equation (3.10) we get

$$\left[1 + \frac{3b + \frac{2\lambda^2}{\mu^2}\tau_{zr}^2}{L^2}\right](\tau_{zr}) = b\mu \frac{\partial v_z}{\partial r} \tag{3.11}$$

where $tr\,\boldsymbol{\tau} = \tau_{zz} = \frac{2\lambda}{b\mu}\tau_{zr}^2$ is used.

Now, we look into the momentum equation (3.5) and integrate it once with respect to '$r$'; and apply the symmetry condition (i.e., at $r = 0$, $\frac{\partial \tau_{zr}}{\partial r} = 0$) at the channel centreline to get $\tau_{zr} = \left(\frac{dp}{dz}\right)\frac{r}{2}$. Substituting this expression of $\tau_{zr}$ in equation (3.11), we get

$$\frac{\partial v_z}{\partial r} = \frac{r}{2\mu}\left(\frac{dp}{dz}\right) + \frac{\lambda^2 r^3}{4b^2\mu^3 L^2}\left(\frac{dp}{dz}\right)^3 \tag{3.12}$$

Integrating equation (3.12) once with respect to '$r$' and applying the no-slip boundary condition results (i.e., at $r = a(z)$, $v_z = 0$)

$$v_z = -\frac{1}{4\mu}\left(\frac{dp}{dz}\right)[\{a(z)\}^2 - r^2] - \frac{\lambda^2}{16\,b^2\mu^3 L^2}\left(\frac{dp}{dz}\right)^3 [\{a(z)\}^4 - r^4] \tag{3.13}$$

which represents the velocity distribution for the pressure-driven flow of a viscoelastic fluid through a deformable cylindrical microchannel. This velocity distribution involves the deformed radius $a(z)$ which is a function of the pressure distribution, i.e., $a(z) = a_0\left\{1 + \frac{p(z)}{2\,G}\right\}$. Since the deformed radius $a(z)$ varies in the axial direction, the velocity component $v_z$ becomes two-dimensional, i.e., function of both '$r$' and '$z$'.

Now, substituting the limit of $\lambda \to 0$, equation (3.13) simplifies to the form $v_z = -\frac{1}{4\mu}\left(\frac{dp}{dz}\right)[\{a(z)\}^2 - r^2]$ which is the velocity distribution for the pressure-driven flow of a Newtonian fluid through a deformable cylindrical microchannel. In absence of any deformation, $a(z) = a_0$; the simplified form of the velocity distribution becomes $v_z = -\frac{1}{4\mu}\left(\frac{dp}{dz}\right)[a_0^2 - r^2]$ which is the well-known expression of the velocity distribution for the pressure-driven flow of a Newtonian fluid through a non-deformable cylindrical microchannel. (Bird RB, Stewart We 2002)

We now calculate the volumetric flow rate ($q$) through the microchannel by integrating the flow velocity (3.13) across the channel cross-section as

$$q = 2\pi \int_0^{a(z)} v_z\, r\, dr = -\frac{\pi a_0^4}{8\mu}\left(\frac{dp}{dz}\right)\left\{1 + \frac{p(z)}{2G}\right\}^4 - \frac{\pi \lambda^2 a_0^6}{24 b^2 \mu^3 L^2}\left(\frac{dp}{dz}\right)^3 \left\{1 + \frac{p(z)}{2G}\right\}^6 \quad (3.14)$$

where the relationship between the deformed and undeformed radius (i.e., $a(z) = a_0\left\{1 + \frac{p(z)}{2G}\right\}$) is substituted. Equation (3.14) involves a term $\left(\frac{dp}{dz}\right)^3$ which makes it non-linear and hence, we cannot get a closed-form expression for the pressure distribution $p(z)$. Thus, we solve equation (3.14) numerically to obtain the pressure distribution. For obtaining the numerical solution, we use the finite element framework of COMSOL Multiphysics. Now the remaining step is to apply the boundary condition for the pressure profile. The boundary condition used here is the outlet pressure at the channel end, i.e., at $z = l$, $p(l) = p_0$. It is important to mention here that the outlet pressure $p_0$ is non-zero because of the presence of the drainage tube at the end of the channel outlet. For calculating pressure $p_0$, one needs to use the Poiseuille's law for the flow of viscoelastic fluid through a rigid (i.e., non-deformable) tube as follows

$$q = -\frac{\pi a_d^4}{8\mu}\left(\frac{dp}{dz}\right) - \frac{\pi \lambda^2 a_d^6}{24 b^2 \mu^3 L^2}\left(\frac{dp}{dz}\right)^3 \quad (3.15)$$

with $a_d$ being the radius of the drainage tube. Equation (3.15) is a cubic equation which has three roots. Two of these roots are complex conjugate to each other and discarded. The remaining real root is considered further. Based on this, we arrive at the following expression of $p_0$

$$p_0 = \left(-\frac{1}{6}\frac{12^{1/3}\left(\alpha_1 12^{1/3}\alpha_2 - \left((\sqrt{3}\sqrt{\alpha_3} + 9q)\alpha_2^2\right)^{2/3}\right)}{\alpha_2\left((\sqrt{3}\sqrt{\alpha_3} + 9q)\alpha_2^2\right)^{1/3}}\right) l_d \quad (3.16)$$

where $l_d$ is the length of the drainage tube. Equation (3.16) involves three coefficients ($\alpha_1$, $\alpha_2$, $\alpha_3$) which are given by $\alpha_1 = \frac{\pi a_d^4}{8\mu}$, $\alpha_2 = \frac{\pi \lambda^2 a_d^6}{24 b^2 \mu^3 L^2}$, and $\alpha_3 = \frac{27q^2\alpha_2 + 4\alpha_1^3}{\alpha_2}$, respectively. Finally, we use this expression of $p_0$ (as written in Equation (3.16)) as the boundary condition at the channel outlet and solve equation (3.14) numerically to determine the pressure distribution. Once $p(z)$ is obtained, the radial deformation $\alpha_r$ at the channel inlet can be calculated as

$$\alpha_r = u_r(a_0, 0) = \frac{a_0}{2G} p(0) \quad (3.17)$$

## 4. Results and discussion

Figure 2 shows the radial deformation patterns during flow of Newtonian liquid (DI water) through microchannels of varying base-to-crosslinker ratios, as commonly encountered in reported microfluidic experiments (Laha *et al.* 2024a; Raj *et al.* 2018; Raj *et al.* 2017). The 10:1 ratio denotes a rigid channel and hence, negligible deformation has been observed over a range of flow rates, while the 30:1 channel, being the softest of the three, records the maximum deformation. It is to be observed that for the 20:1 channel, deformation is rather low for smaller flow rates (< 200 μL/mins), while for the 30:1 channel, significant deformation values have been observed over the entire flow rate range, hence, beyond this, for all our experimental runs we have resorted to the 30:1 channel composition.

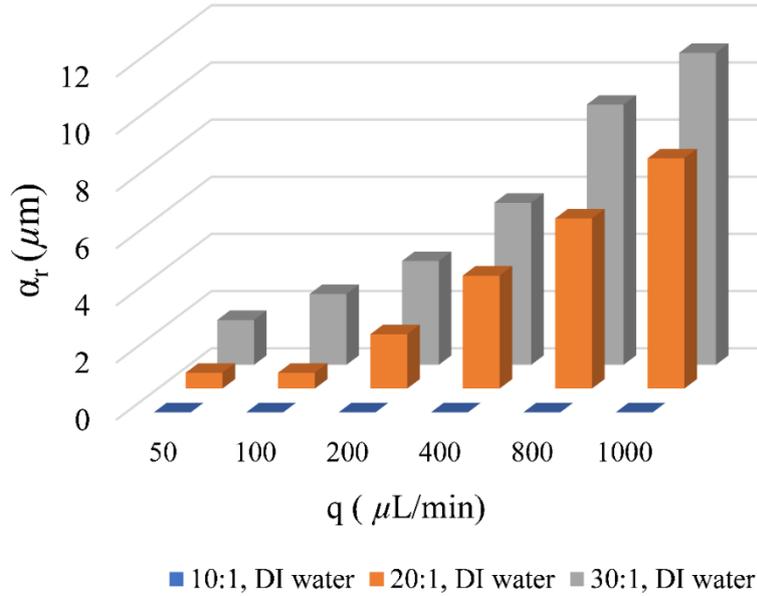

**Figure 2.** Inlet deformation for different elastomer to crosslinker ratio mixtures. The highest channel deformation has been obtained for the 30:1 mixture across all operating flow rates, while the 10:1 mixture produces a typically rigid channel with negligible deformation. DI water is used as the operating fluid.

In Figure 3(a) and Figure 3(b), we have highlighted the estimates of channel inlet deformation as functions of polymer concentration and polymer molecular weight respectively. Before discussing the results, we present a brief overview of the viscoelastic fluid rheology. It is important to mention that the polymer concentration is a very crucial parameter depending on which there is a significant alteration in fluid viscosity and relaxation time (Del Giudice *et al.* 2015). Since the degree of viscoelasticity is decided by the relative strength of elastic and viscous effect, altering the polymer concentration can lead to a transition from one regime to another (Del Giudice *et al.* 2015; Mukherjee *et al.* 2017). For example, if the polymer concentration exceeds a threshold value (which is termed as the cross-over or overlap concentration $c^*$), the interaction between the polymer chains significantly influences the fluid rheological properties (R.G.Larson 1999). Researchers have demarcated three distinct regimes based on concentration, namely, (i) dilute, (ii) semi-dilute unentangled, and (iii) semi-dilute entangled (Del Giudice *et al.* 2015). In this study, we have chosen the range of concentrations in such a way that it covers both dilute ($0.5c^*$ and $c^*$) as well as semi-dilute regimes ($2c^*$ and $4c^*$).

Figure 3(a) highlights the dependence of the inlet deformation ($\alpha_r$) on the polymer concentration for the flow rate of 1000 $\mu$L/min. Inset shows the variation of the same for flow rate of 50 $\mu$L/min. For high flow rate (1000 $\mu$L/min), $\alpha_r$ vs. concentration relation is linear up to $c^*$, beyond which non-linearity sets in bringing about a sharp change in the dependence with concentration, as demarcated by two distinct colour maps in Figure 3(a). This phenomenon is direct evidence of the impact of polymer concentration on the strengthened viscoelastic effect governed by the polymer chain interactions in semi-dilute regimes. For low flow rate (50 $\mu$L/min), $\alpha_r$ vs. concentration relation remains almost constant up to concentration $2c^*$ beyond which deformation increases with polymer concentration. This highlights the fact that the degree of interaction between polymer chains is strongly dependant on the extent of perturbation imparted by the surrounding fluid, which is directly dictated by the imposed flow rate.

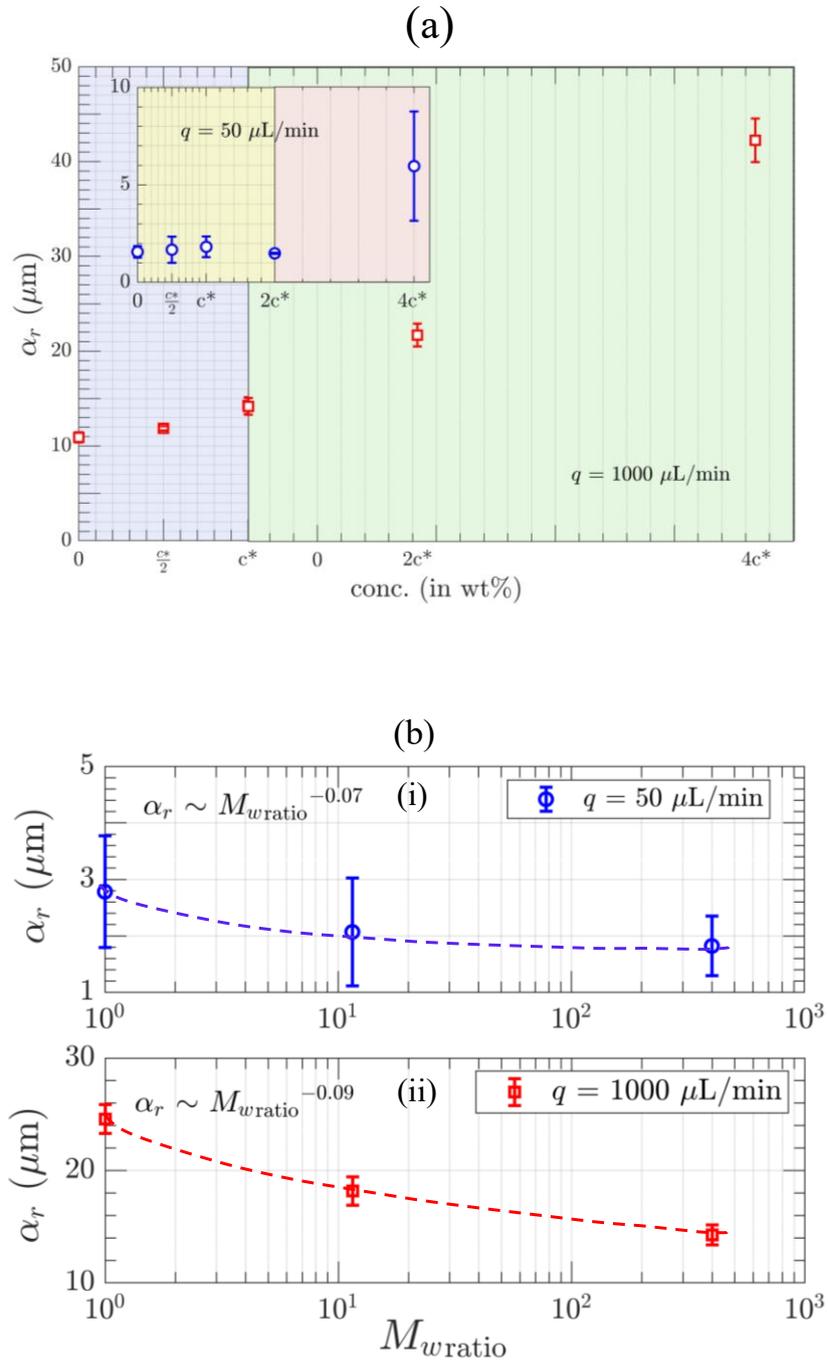

**Figure 3.** (a) The dependence of the experimentally measured channel inlet deformation on the polymer concentration for two different flow rate conditions. Here, we have used aqueous polymer solutions of PEO of molecular weight: 4000000 g/mol. (b) The influence of the polymer molecular weight on the channel inlet deformation. Here, we have used three different polymers PEG (molecular weight: 10000 g/mol), PVA (molecular weight: 115000 g/mol), and PEO (molecular weight: 4000000 g/mol), respectively all at cross-over concentration c*. All experimental data reported have been repeated three times and presented with error bars in order to account for the reproducibility of the trend.

Next, we shift our focus towards the effect of polymer molecular weight on the estimates of inlet deformation, as highlighted in Figure 3(b). For a high molecular weight polymer like PEO, addition of even a small amount ($c^* = 0.071$ % w/vol) of polymer in Newtonian solvent initiates the transition from dilute to semi-dilute regime. On the contrary, for both PVA and PEG, significantly high proportions ($c^* = 1.06$ % w/vol and $c^* = 5.3$ % w/vol, respectively) of polymer needs to be added. Additionally, the fluid relaxation time of PEO is several orders of magnitude higher compared to that of PEG and PVA (Varma *et al.* 2020). This makes the fluid elasticity negligible for low molecular weight polymers. Considering the molecular weight of PEG as a reference scale, a parameter named molecular weight ratio ($Mw_{ratio}$ = Polymer molecular weight / molecular weight of PEG) is defined with respect to which the variation of the inlet deformation is examined for two different flow rates 50 $\mu$L/min. and 1000 $\mu$L/min, respectively. In all the cases, polymer solutions are at cross-over concentration $c^*$. The pronounced viscous effects resulting from higher cross-over concentration of low molecular weight polymers, leads to significantly higher inlet deformation as compared to the case of PEO. As shown in Figure 3(b), the inlet deformation $\alpha_r$ decays with the molecular weight ratio ($Mw_{ratio}$) following a power-law decay: $\alpha_r \sim Mw_{ratio}^{-n}$ (where $n$ is the coefficient of power-law decay) type behavior. Comparing figures 3(b) (i)-(ii), it is evident that the degree of reduction of $\alpha_r$ with $Mw_{ratio}$ enhances upon increasing the flow rate as can be checked by comparing the exponents.

Figure 4(a) highlights the experimental as well as theoretical estimates of the inlet deformation of the channel as a function of volumetric flow rate using different concentrations of aqueous polymer solutions. A reasonably good agreement is found between the experimental measurements and theoretical predictions. Inset shows the results using DI water. With increasing polymer concentration, the elastic effect becomes more pronounced. For polymer solutions in dilute regime (i.e., the concentration is less than equal to overlap concentration $c^*$), the elastic stress introduces a perturbation to the flow field which is propagated to the solid domain through the normal stress balance at the fluid-solid interface. Thus, by altering the polymer concentration, both the hydrodynamics as well as the deformation characteristics get influenced. There is up to ~ 1.6 times increment in the inlet deformation occurring at flow rate 1000 $\mu$L/min when we move from DI water to aqueous PEO solution of concentration $c^*$. When we move from dilute to semi-dilute regime (i.e., the concentration is greater than overlap concentration $c^*$), the polymer chain starts to interact with each other. This strongly influences the rheological properties; fluid viscosity and relaxation time now become strong functions of polymer concentration (Del Giudice *et al.* 2015). This interaction between polymer chains introduces an additional source of perturbation to the flow field and thus leads to larger estimates in inlet deformation, particularly amplified at high flow rates. For example, the inlet deformation is about ~ 2.9 times for concentration $4c^*$ as compared to that for $c^*$ concentration at flow rate 1000 $\mu$L/min.

Figure 4(b) shows the corresponding pressure-drop vs flow rate relationship for varying polymer concentrations. The strengthened elastic stress with increasing polymer concentration gives rise to a strong perturbation to the flow field, thereby resulting in significant augmentation in pressure drop as we move from dilute to semi-dilute regime. Apart from that, the degree of shear-thinning behavior of the solution also increases with increasing polymer concentration which results in reduction of the viscous resistance, leading to higher pressure drop (Varma *et al.* 2022) . As can be seen from figure 4(b), the pressure-drop vs. flow rate relationship is weakly non-linear for dilute solutions and after which the degree of non-linearity increases when we move from dilute to semi-dilute regime. Although, the pressure-drop vs. flow rate curve for $4c^*$ concentration is highly nonlinear up to flow rate of ~ 600 $\mu$L/min, it reaches a saturation at higher flow rates. Recalling the scaling relationships of fluid viscosity ($\mu$) and relaxation time ($\lambda$) with polymer concentration in semi-dilute regime: $\mu \propto c^{1.54}$ and $\lambda \propto c^{0.54}$ (Del Giudice *et al.* 2015), one can understand that, the rate of increment in $\mu$ with concentration is much higher compared to that occurring for $\lambda$. Thus, the observed saturation is attributable to a competition between the enhanced shear-viscosity and the elastic stress-governed perturbation at high concertation.

Finally, figure 4(c)-(d) depicts the deformation and pressure drop characteristics by deploying polymers of different molecular weight. Figure 4(c) shows a comparison of the inlet deformation estimates with varying flow rate for two different polymer solutions PEO and PEG. Owing to the much higher cross-over concentration $c^*$, the shear viscosity of aqueous PEG solution is significantly higher compared to that of PEO solution at $c^*$. This is primarily responsible for an enhanced degree of inlet deformation compared to that of PEO as depicted in Figure 4(c), despite the fluid relaxation time of the former being several orders of magnitude lower. It is interesting to note that deploying PEG solution of concentration $c^*$ is almost analogous to deploying a Newtonian fluid of higher viscosity compared to DI water. The associated pressure-drop vs. flow rate characteristics is shown in Figure 4(d) where the enhanced fluid viscosity for PEG gives rise to an enhancement in the magnitude of pressure drop as well as in the degree of non-linearity of the pressure-drop vs. flow rate relationship.

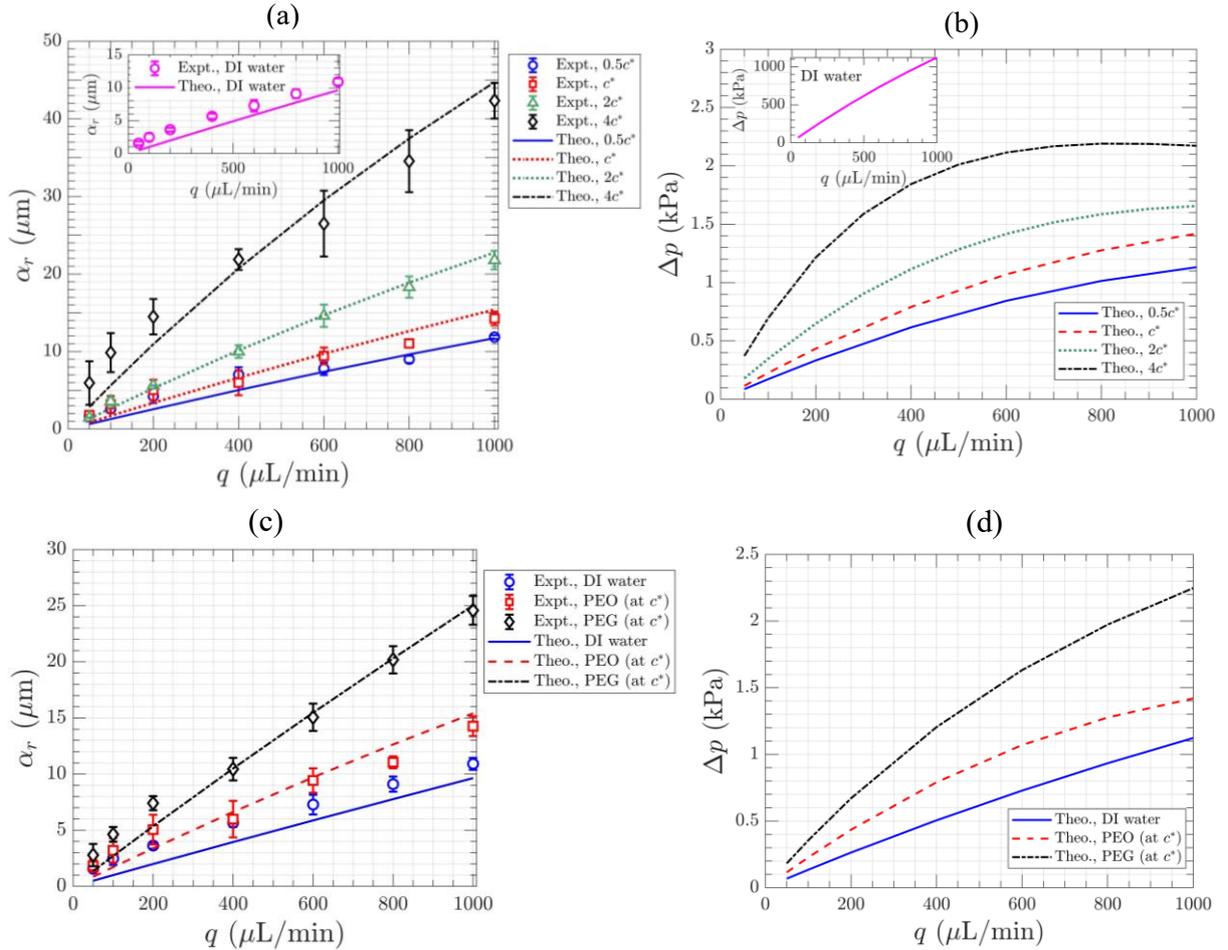

**Figure 4.** (a) The comparison between the experimental (solid-coloured markers) and theoretical (lines) inlet deformation values for aqueous PEO solutions of different concentrations in dilute and semi-dilute regime; (b) the pressure-drop vs flow rate characteristics for different concentrations of aqueous PEO solutions; (c) the inlet deformation vs flow rate plot for two different polymer suspensions PEO and PEG having significantly different molecular weights. (d) shows the pressure drop vs flow rate characteristics for the two different polymer suspensions. Results using different polymer suspensions in both (c) and (d) have been benchmarked against their corresponding Newtonian counterparts.

## 5. Conclusions

We reported controlled experiments on the elasto-mechanical behaviour of compliant microchannels carrying aqueous polymer solutions having viscoelastic rheology. Our experimental results revealed how the inlet deformation varies with flow rate for dilute as well as semi-dilute solutions. The rate of increase in inlet deformation with polymer concentration showed a sharp change in slope beyond a cross-over PEO concentration, emphasizing strong dependence on the polymer chain interactions. At the same crossover concentration, a solution of low molecular weight polymer (approximately two orders of magnitude lower) exhibited a weakly non-linear deformation of considerably higher magnitude. This could be attributed to the substantially reduced elastic contributions to the polymer stress, coupled with an increased viscous effect resulting from the elevated polymer concentration beyond the dilute regime. Theoretical insights revealed the implications of a nonlinear interplay between the fluid rheology and the channel deformation behaviour, closely corroborating with the experimental findings.

While our results provided valuable physical insights into the role of polymer molecular weight and concentration on the dynamic response to flow in deformable microchannels, these findings remain relevant only up to a certain threshold flow rate. At higher flow rates, viscoelastic fluids can exhibit instabilities that may be intensified by the deformability of the channel, leading to complex and sometimes chaotic flow patterns (Chandra *et al.* 2019). In such cases, the entry and exit regions of the deformable microchannel may experience significant changes in flow behaviour due to sudden wall deformation. The elasticity of both the fluid and the channel walls can result in phenomena such as stress overshoots or the formation of viscoelastic boundary layers (Escott & Griffiths 2023). However, these effects were not explored in our work, as we focused on regimes of low Reynolds number flow.

The practical implications of the results reported in this work are extensive and impactful. Deformable microchannels, for example, can simulate the small pores found in reservoir rocks, enabling researchers to explore how complex fluids move through these confined spaces (Wang *et al.* 2010). By investigating the behaviour of viscoelastic fluids in deformable channels, engineers can enhance oil recovery methods, such as optimizing polymer flooding - where increasing the viscosity of injected water boosts oil extraction efficiency (Speight 2013; Zeynalli *et al.* 2023). Understanding flow in deformable channels allows for more accurate reservoir simulations, improving predictions of how various recovery strategies will perform in real-world conditions. In polymer processing, insights gained from studying flow through deformable microchannels are critical for advancing techniques that lead to better material control, more efficient manufacturing processes, and higher-quality products (Cheng *et al.* 2022). These channels reveal how changes in pressure, temperature, and geometry affect polymer flow on a microscale, which is crucial for energy conversion, biomedical microdevices and flexible electronics (Li *et al.* 2014; Nguyen *et al.* 2013; Roy *et al.* 2024). Understanding polymer flow in these contexts is essential for designing moulds and processes that produce high-precision microstructures reliably. Deformable microchannels also play a central role in the development of advanced wearable technologies, influencing their functionality, reliability, comfort, and adaptability(Annabestani *et al.* 2022). Flexible electronics in wearables, such as fitness trackers and medical sensors, rely on conductive polymers flowing through deformable microchannels to maintain the designed electrical connectivity even when the devices are stretched, bent, or twisted(Guo *et al.* 2021; Kim *et al.* 2018). Studying this flow ensures reliable performance in wearable electronics and aids in the design of soft robotics components that enhance movement and tactile feedback in these devices (Biswas *et al.* 2020; Gao & Ren 2021; Gao *et al.* 2016; Katchman *et al.* 2018; Laha *et al.* 2022, 2023; Laha & Chakraborty 2023). Moreover, controlling fluid flow in deformable microchannels opens up innovative possibilities for wearables, such as adaptive textiles and dynamic displays embedded in clothing(Xing *et al.* 2023). This understanding enables the creation of more advanced, multifunctional wearable devices. Additionally, deformable microchannels are fundamental to developing microvasculature-on-a-chip technologies (Benam *et al.* 2016; Bhattacharya *et al.* 2022; Huh 2015; Priyadarshani *et al.* 2021). These in-vitro models may simulate the elasticity and mechanical properties of blood vessels to a large extent, including unleashing how vessels respond to varying pressures and flow rates. Such models are thus invaluable for studying physiological processes, drug interactions, and the effects of flow stresses on vascular health (Berry *et al.* 2021; Luque-González *et al.* 2020; Shakeri *et al.* 2023; Wang *et al.* 2023). They offer a more realistic assessment of drug dynamics compared to traditional pre-clinical trials and can be tailored to individual patients, leading to more precise and effective therapies (Chopra *et al.* 2023). The ability to simulate patient-specific vascular characteristics in these systems represents a significant advancement over current empirical approaches, providing a more accurate and personalized basis for treatment decisions than ever before (Palasantzas *et al.* 2023).

**Funding statement.**

**Declaration of interests.** The authors declare no conflict of interest.

**Author contributions.** S.L., S.M. and S.C. designed the research problem. S.L. performed experiments and image analysis. S.M. performed theoretical calculations and rheometry measurements. S.C. acquired funding and supervised the research. S.L., S.M. and S.C. analysed the data and prepared the manuscript.

**Data availability statement.** All data, calculation and derivations are included in the article and supplementary material.

# Supplementary Material

# Dynamical Response of Deformable Microchannels under Pressure-Driven Flow of Aqueous Polymer Solutions

Sampad Laha[1], Siddhartha Mukherjee[2], and Suman Chakraborty[1]*

[1]Department of Mechanical Engineering, Indian Institute of Technology Kharagpur, Kharagpur, West Bengal, 721302, India.

[2]Department of Chemical Engineering, Indian Institute of Technology Guwahati, Guwahati, Assam, 781039, India.

*Email: suman@mech.iitkgp.ac.in

**Table S.1: Youngs's modulus for different base to crosslinker ratio of PDMS**

| PDMS Base: Crosslinker | Young's Modulus (MPa) |
|---|---|
| 10:1 | 2.81  (Raj *et al.*, 2018) |
| 20:1 | 1.57  (Fincan 2015) |
| 30:1 | 0.157  (Raj *et al.* 2017) |

**Table S.2: Rheological parameters of different concentrations of aqueous polyethylene oxide (PEO) solution**

| concentration | wt/vol % | $\mu$ (in Pa.s.) | $\lambda$ (ms) |
|---|---|---|---|
| $c*/2$ | 0.035 | 0.0013 | 1.052 |
| $c$ | 0.071 | 0.0017 | 1.052 |
| $2c*$ | 0.142 | 0.0027 | 1.528 |
| $4c*$ | 0.284 | 0.0058 | 2.219 |

For aqueous PEO solutions, the relaxation time for dilute solutions (i.e., concentration less than or equal to $c*$) are calculated using Zimm's definition ($\lambda_z$), i.e., $\lambda_z = \frac{1}{\zeta(3\nu)} \frac{[\eta] M_w \eta_s}{N_A k_B T}$ (Tirtaatmadja *et al.* 2006) in which $M_w$ is the polymer molecular weight, $\frac{1}{\zeta(3\nu)}$ front factor, $[\eta]$ intrinsic viscosity, $\eta_s$ solvent viscosity, $T$ absolute temperature, $k_B$ Boltzmann constant, and $N_A$ Avogadro's number. The intrinsic viscosity is calculated using the Mark-Houwink-Sakurada (MHS) equation $[\eta] = 0.072 M_w^{0.65}$ (Tirtaatmadja *et al.* 2006). Once $[\eta]$ is calculated, we determine the cross-over/overlap concentration c* using the relation $c^* = \frac{1}{[\eta]}$. Knowing the value of c*, both dilute as well as semi-dilute solutions are prepared. For solutions in semi-dilute regime, the relation time of fluid is calculated using the scaling relation $\lambda_{semi-dilute} \propto c^{0.54}$ (Del Giudice et al. 2015) where c is the polymer concentration. Apart from that, two solutions are prepared using aqueous polyvinyl alcohol (PVA) and polyethylene glycol (PEG) solutions for which the overlap concentrations (c*) are 1.06 % w/vol and 5.3 % w/vol, respectively. Zimm's relaxation times ($\lambda_z$) are 2 x 10$^{-3}$ ms and 30 x 10$^{-6}$ ms, respectively (Varma et al. 2020).

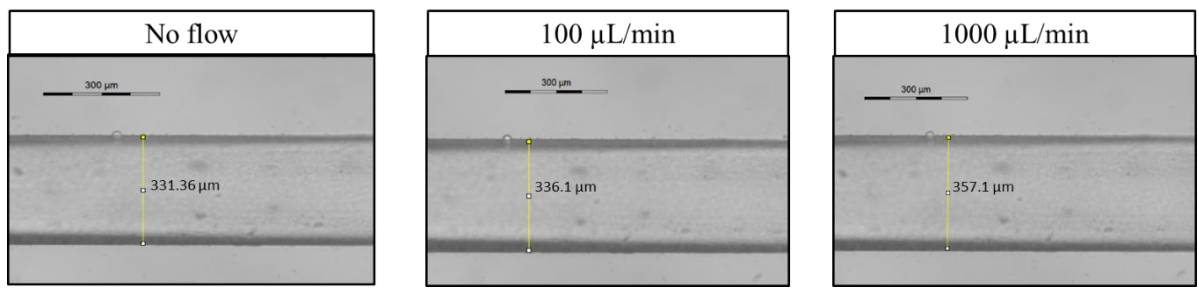

**Fig. S1**: Microscopic images showing microchannel inlet deformation profile at flow rates of zero, 100 µL/min and 1000 µL/min (left to right) with PEO solution at c* concentration.